\documentclass[reprint,superscriptaddress,prl]{revtex4-1}

\usepackage{graphicx}
\usepackage{amssymb}
\usepackage{amsmath}
\usepackage{epsfig}
\usepackage{chemarr}
\usepackage{url}
\usepackage{natbib}
\usepackage{dcolumn}
\usepackage{bm}
\usepackage{color}
\usepackage{ulem}

\def\be{\begin{equation}}
\def\ee{\end{equation}}
\def\bmu{\begin{multline}}
\def\bea{\begin{eqnarray}}
\def\eea{\end{eqnarray}}

\def\p{\partial} 
 
\def\nn{\nonumber}
\def\f{\frac}

\def\l{\left(}
\def\r{\right)}

\def\grad{{\bf \nabla}}
\def\v{{\bf v}}
\def\a{\alpha}

\newcommand{\me}{\mathrm{e}}

\begin{document}

\title{Robust Molecular Computation by Active Mechanics}
\author{Kabir Husain}
\affiliation{Department of Physics, University of Chicago, Chicago, USA}
\author{Sriram Ramaswamy}
\affiliation{Centre for Condensed Matter Theory, Department of Physics, Indian Institute of Science, Bangalore 560012, India}
\author{Madan Rao}
\affiliation{Simons Centre for the Study of Living Machines, National Centre for Biological Sciences - TIFR, Bangalore 560065, India}
\date{\today}

\begin{abstract}
The living cell expends energetic and material resources to reliably process information from its environment. To do so, it utilises unreliable molecular circuitry that is subject to thermal and other fluctuations. Here, we argue that active, physical processes can 
provide error correcting mechanisms for information processing. We analyse a model in which fluctuating receptor activation induces contractile stresses that recruit further receptors, dynamically controlling resource usage and accuracy. We show that this active scheme can outperform passive, static clusters (as formed, for instance, by protein crosslinking). We consider simple binary environments, informative decision trees, and chemical computations; in each case, active stresses serve to contextually build signalling platforms that dynamically suppress error and allows for robust cellular computation. 
\end{abstract}

\maketitle


%

Living cells process external information to respond to a dynamic and informative environment: in other words, they 
compute. Their computational hardware is comprised of assemblies of specialised proteins~\cite{Bray1995} whose spatiotemporal organisation is regulated in an ATP-dependent manner -- from dynamic clusters of receptors at the cell surface~\cite{Goswami2008,VanZanten2009,Kalappurakkal2020} to transient transcriptional factories in the nucleus~\cite{Hnisz2017,Bialek2019}. While studies of information flow in cellular systems have focused on biochemical processes~\cite{Bialek2012,Phillips2020}, the role of cellular mechanics in regulating and enabling such computations remains less appreciated~\cite{Chaudhuri2011}.

The natural framework to describe the mechanical regulation of cellular material is active mechanics~\cite{Marchetti2013}, which describes the physical organisation and dynamics that result from non-equilibrium stresses. In contrast, the study of cellular information processing is rooted in the language of biochemistry, Markov processes, and information theory, which together describe how ligand binding, conformational changes, and post-translational modifications conspire to transduce environmental information into cellular responses \cite{Bialek2012,Tkacik2016,Phillips2020}. Yet, these processes are intimately coupled: 
on the one hand, non-equilibrium stresses in the cell are tightly regulated by upstream signalling pathways~\cite{Levayer2012}. On the other hand, the spatiotemporal organisation of many of these signalling pathways is driven by nonequilibrium stresses~\cite{Kalappurakkal2020}.

This interplay between active mechanical stresses and chemical signalling must form the basis for molecular information processing and chemical computation in the cell. 
Examples include the cortical actomyosin-driven dynamic assembly of the components of the cell adhesion \cite{VanZanten2009,Lecuit2015,Kalappurakkal2019} and immune signalling machinery \cite{Kumari2014,Tolar2017} 
at the cell surface, as well as the dynamic organisation of transcriptional machinery at the foci of active transcription \cite{Hnisz2017,Cho2018,Sabari2018,Bialek2019}.

Here, we present a
theoretical proof of principle of the information processing capabilities of such actively driven signalling platforms, whose constituent elements couple internal chemical state changes with non-equilibrium physical stresses. We show that while individual molecular information processing units (IPUs) may be prone to error and unreliable, actively driven clusters can make robust, informed decisions. The performance of the cluster depends on the strength of the active driving. By considering layered, informative environments inspired by real biological problems, we find that the active driving naturally extracts more information from the environment over time.

\begin{figure}
\includegraphics[width=.8\linewidth]{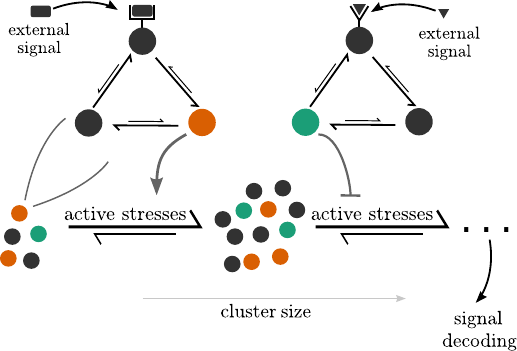}
\caption{ \label{fig:schem} A schema of the non-equilibrium information processing we consider, in which external signals incident on molecular information processing units (IPUs) drive internal chemical state changes that activate or inhibit local mechanical stresses. These mechanical stresses, in turn, lead to transitions in the internal chemical state and recruitment of additional units, thereby increasing the cluster size until final dissolution following the decoding of the signal. }
\end{figure}

We describe the background physical medium as a viscous gel damped by a substrate and endowed with active contractile stresses $\boldsymbol{\sigma}_0$ originating in the actomyosin machinery, with fluctuations $\delta\boldsymbol\sigma$ whose spatiotemporal correlations  
\be
\langle \delta \boldsymbol\sigma({\bf 0},0) \,\delta \boldsymbol\sigma({\bf r}, t) \rangle  = \mbox{\sffamily\bfseries f}
(r/\xi,t/\tau) 
\label{eq:activenoise}
\ee 
vanish rapidly beyond length- and timescales $\xi$ and $\tau$ respectively.
Such active contractile stress fluctuations have been studied in \cite{lau2003microrheology,Basu2008,SinghVishen2018}, and can be dynamically realised using a birth-death process of contractile stress events localised over a region $\xi$ with a lifetime $\tau$ \cite{Chaudhuri2011, Das2016}. 

Embedded in this active medium are molecular components with varied response to active stress fluctuations. Some are {\it passive  scalars}~\cite{Chaudhuri2011,Gowrishankar2012}, simply advected by contractile flows generated by the active stress fluctuations. Others, such as the cellular proteins integrin \cite{Kalappurakkal2019} and cadherin \cite{Lecuit2015} and immune receptors \cite{Kumari2014}, can trigger activation of the actomyosin machinery to locally enhance contractility, while molecules such as tropomyosin or phosphatases that shut down signalling~\cite{Levayer2012} can trigger repression of actomyosin activity, leading to a dissolution of preexisting active stresses. Such stress activators and inhibitors \cite{Chaudhuri2011,Bois2011,Kumar2014} are {\it active scalars} -- which locally modify the background active stress from $\boldsymbol \sigma_0$ to 
\be
\boldsymbol\sigma =   \boldsymbol{\sigma}_0 - \zeta \Delta\mu \sum_\a c_\a \, \beta_\a \,\mathbb{I}
\label{eq:activestress}
\ee
where $c_{\a}$ is the non-dimensional concentration of species $\a$, and $\beta_{\a}$ is positive, negative or zero for activators, inhibitors or passive components respectively~\cite{Chaudhuri2011,Kumar2014}, $\mathbb{I}$ is the unit tensor, and we have made explicit the familiar \cite{Marchetti2013} product of a cross-kinetic coefficient $\zeta$ (negative for contractile stress) and chemical driving force $\Delta \mu$, to which both $\boldsymbol\sigma_0$ and its modification are proportional. Note that we have in mind here isotropic contractility in a layer such as the actin cortex.

Crucially, particles moving in the active medium can switch (Fig.~\ref{fig:schem}) between the chemical states labelled by $\a$, thus altering their interaction with the embedding active medium. For instance, ligand binding to a receptor in the cell membrane may switch it from being passively advected by cortical flows to actively remodelling cortical actomyosin.
Putting these features together, the concentrations $c_\a$ evolve via advection by the gel velocity $\mathbf{v}$,
gradients in chemical potential $\mu_\a$,
and interconversions between species with rates $k_{\a \to \beta}$: 
\bea 
\label{eq:conc}
\f{\p}{\p t}c_{\a} &=& -\grad \cdot \l c_\a \mathbf{v} \r + M_\a \nabla^2 \mu_\a \nn \\ &+&
\sum_{\beta \neq \alpha} \, k_{\beta \to \a}(\boldsymbol \sigma) c_{\beta} - k_{\a \to \beta}(\boldsymbol \sigma)c_{\a} 
\eea
where $M_\a$ is the mobility of the $\a$th component~\footnote{The currents of active model B and B+ \cite{tjhung2018cluster,cates2018theories} enter at higher order in gradients than the effects considered in \eqref{eq:activestress}-\eqref{eq:velphi}.}.
The velocity $\v$ is determined by local force balance
\be
 \label{eq:velphi}
    \gamma \v =   \eta \nabla^2 \v +  \nabla  \cdot \boldsymbol \sigma
 \ee
where ${\boldsymbol \sigma}$ is given by \eqref{eq:activestress}
and we have taken the system to be in contact with a substrate characterized by a frictional damping constant $\gamma$.
One non-reciprocal aspect of the dynamics \eqref{eq:conc} is worth noting: $\v$ advects both passive and active species, but only active species contribute to the stress that drives $\v$.


The dynamics of clustering in such a medium thus depends both on an internal chemical coordinate and a collective mechanical coordinate (e.g. stress). It is this coupling between chemical transitions and nonequilibrium flows that integrates information from the constituent molecular information processing units in a cluster, Fig.~\ref{fig:schem}.

\indent
{\it Master equation for the active cluster:} 
Defining the natural velocity scale $U=\vert \zeta \Delta \mu\vert/\sqrt{\eta \gamma}$ \cite{Kumar2014}, the P\'{e}clet ($\mbox{Pe}= U\xi/D$) and Damk\"{o}hler ($\mbox{Da}= {k\xi^2/D}$) numbers characterize respectively the importance of advection and chemical reaction relative to diffusion, on the scale of the correlation length $\xi$, and thus describe the chemical response to a physical driving~\cite{bandopadhyay2017enhanced}.
Here we will work in the regime $Pe > Da$; because of the contractile nature of the active stresses, the enhancement of chemical reactions will occur in spatially localised hotspots of reactivity~\cite{Chaudhuri2011}.

We study the compositional dynamics of one such hotspot (a `cluster' of size $\xi$). Eq. \eqref{eq:conc} implies that the populations $n_\a \equiv \int_{\xi} c_\a({\bf r})$ can change through advection, diffusion and reaction. The result is a master equation for the joint probability $P(\ldots, n_\a,\ldots)$ that there are $n_1$, $n_2$, $\ldots, n_\a,\ldots$ particles of types $1$, $2$, ..., $\a$... in the cluster:
\begin{widetext}
\bea \label{eq:bigoldmastereqn}
\f{\p}{\p t} P(\ldots, n_\a,\ldots) = && \sum_\a  \l k_+ P(\ldots, n_\a-1,\ldots)  - k_+ P(\ldots, n_\a,\ldots) + k (n_\a + 1) P(\ldots, n_\a+1,\ldots) - k n_\a P(\ldots, n_\a,\ldots) \r \nn \\
&& + \sum_{\alpha} \sum_{\beta \neq \alpha} k_{\beta \to \a} \l \l n_{\beta} + 1 \r P(\ldots, n_\a-1,n_{\beta} + 1,\ldots) - n_{\beta} P(\ldots, n_\a,\ldots) \r \nn \\
&& + \sum_{\alpha} \sum_{\beta \neq \alpha} k_{\a \to \beta} \l  \l n_\a + 1 \r P(\ldots, n_\a+1,n_{\beta} - 1,\ldots) - n_\a P(\ldots, n_\a,\ldots) \r  
\eea
\end{widetext}

The first four terms in \eqref{eq:bigoldmastereqn} describe the recruitment and loss of particles from the cluster with rates $k_+$ and $k$. These are a consequence of the advective flux in \eqref{eq:conc}, arising from the active stress in \eqref{eq:velphi} which, in turn, depends on the number of active particles in the cluster (i.e. on $\{ n_\a \}$). The rate of recruitment $k_+$ is therefore \textit{state-dependent}: $k_+ = k_+( \,\sigma(\{n_\a\}) \,)$. The remaining terms describe particle interconversions.


Note that the dynamics of cluster growth and chemical composition are thereby coupled -- chemical state transitions affects the active stress $\sigma$, which drives the dynamics of cluster growth. What this means is that external chemical signals such as ligand binding can trigger the formation or dissolution of clusters.

\begin{figure}
\includegraphics[width=\linewidth]{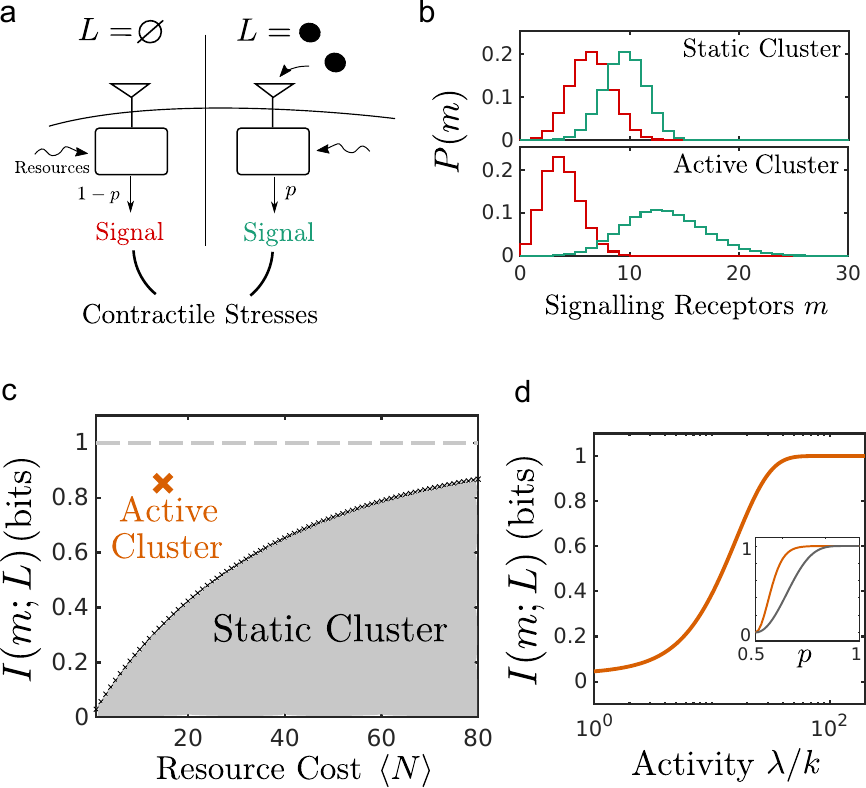}
\caption{ \label{fig:recognition} \textbf{Active clustering allows for efficient molecular information processing}. \textbf{a} A noisy receptor, which signals inappropriately (i.e. in the absence of a ligand) with probability $1-p$. Signalling channels activate contractile stresses, which recruit more receptors to a cluster. 
\textbf{b} The distribution of signalling receptors $m$ in a cluster, with (green) and without (red) ligand, for a static cluster and an active cluster of average size $N=15$. \textbf{c} An active cluster can beat the trade-off, exhibited by static clusters, between information transmission (quantified by the mutual information $I(m;L)$) and resource cost ($\propto N$). \textbf{d} Mutual information as a function of active driving and (inset) as a function of molecular fidelity $p$ (orange: active cluster, grey: static cluster). Unless otherwise noted, $p = 0.6$ and $\lambda/k = 25$.}
\end{figure}





This ability to contextually regulate cluster size and composition endows active fluids with the ability to efficiently process information. To demonstrate this, we define the overall cluster size $n = \sum_{\a} n_\a$ and average Eq.\,\ref{eq:bigoldmastereqn} over the composition $f_\a \equiv n_{\alpha}/n$, assuming that the chemical state transitions $\alpha \to \beta$ are fast compared to cluster growth:
\bea \label{eq:finalmaster}
\f{\p}{\p t} P_n(t) &=& k_+(n-1) P_{n-1} -  k_+(n) P_n  \nn \\  &+&  k \l (n+1) P_{n+1} - n P_{n} \r
\eea
\noindent where the state-dependent rate $k_+$ depends on $n$ implicitly through the active stress $\sigma$. To fix a form for $\sigma(n)$, we suppose that the active stress only takes on discrete values -- such that $k_+ = \lambda \propto \me^{-\zeta \Delta \mu} $ when the active stress is \textit{on} and is $0$ otherwise. The ratio $\lambda/k$ therefore quantifies 
activity relative to diffusion, standing in for the P\'eclet number mentioned above.




\indent
{\it Receptors as information processing units:} To characterise the information transmission properties of an active cluster we couple the internal state transitions $k_{\alpha \to \beta}$ to external ligands. We consider first the case of a simple receptor, which binds to an external ligand $L$ and initiates signalling when the ligand is bound. As molecular events are noisy, receptors may erroneously signal even when the ligand is absent, or fail to signal in the presence of a ligand, with probability $1-p$, Fig.~\ref{fig:recognition}a. The mutual information between ligand presence and receptor signalling is simply that of a binary symmetric channel~\cite{Cover2005},
\be
I_{\text{single}} = 1 + p\log p + (1-p)\log (1-p)
\ee
\noindent where we have assumed that the ligand is equally likely to be present or absent (the least informative environment): $P(L) = 1/2$, and all logarithms are to the base $2$. We consider a single-channel fidelity of $p=0.6$, which results in an information transmission capacity of only $I \approx 0.03$ bits for a single receptor; molecular noise is therefore a severe limitation.

Crucially, we now couple receptor signalling to contractile stresses, such that the active stresses are on when the majority of receptors in a cluster are signalling. Denoting by $m$ the number of signalling receptors in a cluster, we find the steady-state solution $P(m)$ of \eqref{eq:finalmaster} in a two-step procedure detailed in section \textbf{Ligand Recognition} of the SI. First, Eq. \ref{eq:finalmaster} can be cast in matrix form $\p_t \vec{P} = \mathcal{L} \vec{P}$, where the matrix $\mathcal{L}$ is tridiagonal. We solve for the steady state $P(n)$ by numerically diagonalising $\mathcal{L}$, assuming a maximum cluster size of $n=300$, much larger than the typical cluster sizes found at these parameters. We then compute $P(m)$ as $\sum_n P(m\vert n)P(n)$, where $P(m\vert n)$ is the conditional probability of $m$ given $n$, and characterise the information transmission between the ligand and cluster composition by computing the mutual information $I(m;L)$ between ligand presence and receptor signalling, 
\bea
I(m;L) &=& - \sum_m P(m)\log P(m) \nn \\ &+& \sum_{m,L} P(m,L)\log P(m\vert L)
\eea
as well as a `resource cost' proportional to the number of receptors in the cluster.

\indent In Fig.\,\ref{fig:recognition}(b) we compare the distributions $P(m \vert L)$ for a static cluster (fixed $n$) with those of an active one at the same average size ($\langle n \rangle$). Both the active cluster and the static cluster break detailed balance in the chemical transitions of the receptor. However, only the active cluster breaks time-reversal symmetry in the physical dynamics of cluster size. As a consequence, by being able to dynamically regulate its size, the actively coupled system only exploits the enhanced resolving capability of a larger cluster in the presence of an external stimuli. This allows it to beat a trade-off -- between discriminatory power and cluster size -- constraining a static cluster, Fig.\,\ref{fig:recognition}(c). For a calculation of this tradeoff see section \textbf{Ligand Recognition} of the SI.

\indent The performance of the active cluster depends on the strength of the underlying active contractile stresses. In Fig.\,\ref{fig:recognition}(d), we plot the mutual information $I(m;L)$ as a function of activity $\zeta \Delta\mu$: a more informative output requires greater activity. We also explore performance as a function of the molecular fidelity $p$, Fig.\,\ref{fig:recognition}(d) (inset): notably, in contrast to a static cluster, the actively regulated cluster is robust to even low values of molecular fidelity, showing how reliable active computation is possible even with unreliable molecular components. 

\begin{figure}
\includegraphics[width=\linewidth]{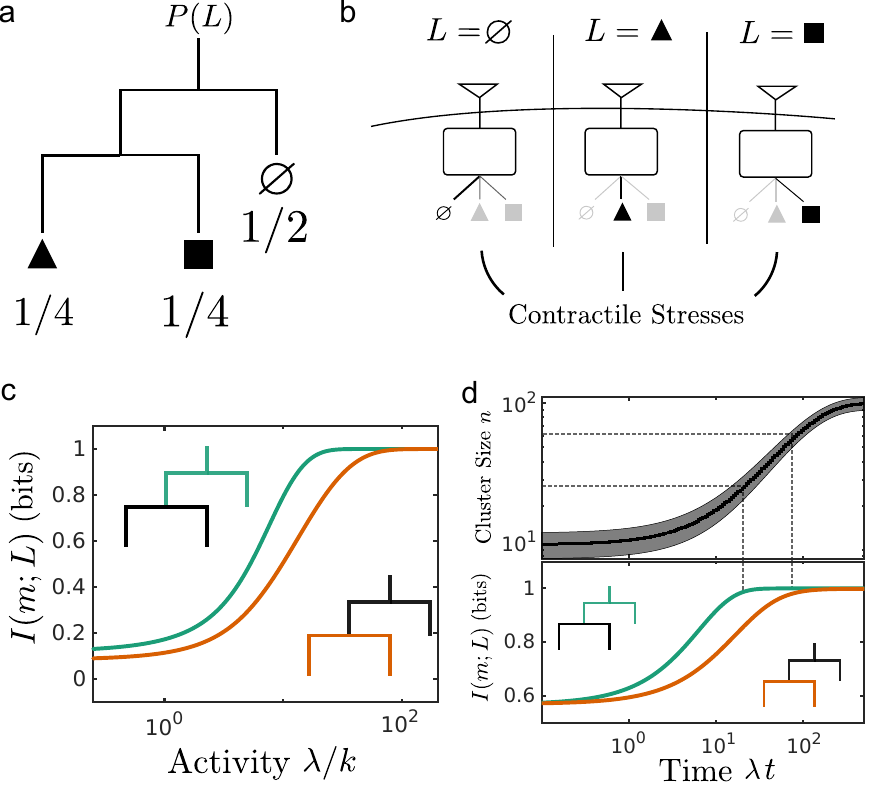}
\caption{ \label{fig:discrimination} \textbf{Discrimination requires higher activity than recognition.} \textbf{a} The environmental states $L$ represented by a decision tree. Ligands may be absent or present; if present, they can be identified as being one of two types (triangle or square). Here, probabilities are assigned as on a uniform binary decision tree. \textbf{b} The receptor, as an information channel, attempts to read the environment. As before, detecting the presence of a ligand actuates contractile stresses that recruit more receptors. \textbf{c} The mutual information as a function of activity, for the ligand recognition problem (green) and for ligand discrimination (orange). Reading deeper into the decision tree, a, requires a stronger active driving as well as \textbf{d} increased cluster size (top) and hence longer times (bottom).}
\end{figure}

In the living cell, recognition of an external ligand typically initiates a sequence of downstream processes that recruit further signalling molecules to extract more information from the environment. We explore this in a hierarchically structured environment $L$, depicted in Fig.\,\ref{fig:discrimination}(a) as a binary decision tree. The ligand may be absent or present; when present, it may be identified as being as one of two types (here shown as a triangle or a square). The task of the information processing machinery is first to recognise the presence of the ligand, and then to discriminate between the two types.

\begin{figure*}
\includegraphics[width=\linewidth]{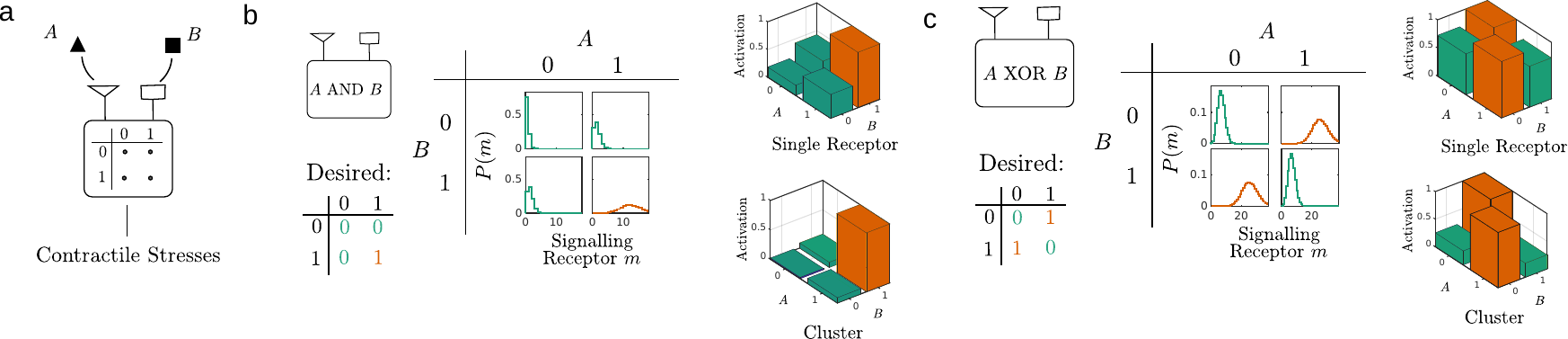}
\caption{ \label{fig:gates} \textbf{Implementing robust molecular logic by active clustering.} \textbf{a} Single receptors read two inputs, $A$ and $B$, and decide to signal based on an encoded truth table. Signalling recruits more receptors via contractile stresses. \textbf{b} and \textbf{c}: Active cluster implementations of molecular AND (\textbf{b}) and XOR (\textbf{c}) gates. Left: the desired truth table. Centre: the distribution of the number of signalling receptors in the cluster, $m$, for different environmental states. Right: the activation (signalling) strength for a single receptor (top) and for an active cluster (bottom, defined as $\langle m \rangle$ and scaled so that maximal activation is $1$.). The active cluster is able to suppress erroneous activation to achieve the desired truth table. Parameters: $p = 0.7$, $\lambda/k = 50$.}
\end{figure*}

\indent This is accomplished by the receptor architecture shown in Fig.\,\ref{fig:discrimination}(b), whose three outputs ($R = 0,1,2$) correspond to the three environmental states ($L = \phi$, triangle, square, respectively). To simultaneously characterise ligand recognition and discrimination, we consider two separate mutual information measures: characterising ligand presence, $I_{\text{recog}}(m_r ; \text{ Ligand present or absent})$, and ligand identity: $I_{\text{discrim}}(m_i ; L = \text{ triangle or square})$, see SI for details.

\indent A large $I_{\text{recog}}$ implies that the cluster can reliably distinguish between ligand presence and absence, independent of ligand type. A large $I_{\text{discrim}}$ indicates a robust discriminatory ability. In Fig.\,\ref{fig:discrimination}(c), we plot these two measures as a function of activity $\zeta \Delta\mu$. Notably, achieving robust discrimination (high $I_{\text{discrim}}$) requires a higher activity than the simpler task of ligand recognition -- extracting more information from the environment requires a stronger coupling to active stresses. From the dynamics, Fig.~\ref{fig:discrimination}(d), we see that the gain of information is \textit{sequential}: larger clusters at later times read further down the decision tree in Fig.~\ref{fig:discrimination}(a).

\indent 
{\it Multiple-input computations:} So far we have looked at examples of single input computations, in which the task is either recognition or discrimination. In certain contexts, the cell is interested in initiating a response to a specific combination of signals. For instance, cell spreading occurs only in the presence of particular chemical (i.e. extra-cellular matrix) and physical (i.e. substrate stiffness) cues. To accomplish this, the cell must perform a logical operation on multiple inputs -- analogous to the logic gates that are foundational in computer science. We consider the receptor architecture in Fig.\,\ref{fig:gates}(a), which accepts two inputs $A$ and $B$ and performs a logical operation $g(A,B)$. The output of the operation is receptor activation ($R=1$) or quiescence ($R= 0$). Once again, we suppose that the receptor is characterised by a fidelity $p$, here corresponding to the correct recognition of each signal $A$ and $B$. 
\indent The probability of activating a single receptor ($R=1$) is shown for the AND and XOR logical operations in the top-right of Fig.\,\ref{fig:gates}(b),(c) (scaled relative to the maximal probability). Regardless of the choice of logical operation $g$, we suppose that receptor signalling ($R=1$) induces contractile stresses and recruits further receptors to the cluster. 

\indent In Fig.\,\ref{fig:gates}(b),(c) we show the distributions $P(m)$ for each value of the inputs $A,B$, for the case of the AND and XOR operations, respectively. In each case, the active cluster correctly achieves states of low and high signalling output $m$ based on the desired truth table. To compare against the case of the single receptor, we took as metric for cluster activation the average $\langle m \rangle$. We show this on the lower-right of Fig.\,\ref{fig:gates}(b),(c). Despite arising from an inherently noisy molecular computation, the output of the cluster robustly reproduces the AND and XOR truth tables. 

\indent
\textit{Discussion:} We have argued that, by controlling their clustering via active contractile stresses, molecular sensing apparatuses can dynamically control their resource usage and fidelity. The information processing capabilities of biochemical networks are often compared against their consumption of free energy, typically in detailed balance breaking chemical dynamics \cite{Lan2012,Sartori2014}. Here, we show instead a role for the breaking of detailed balance in physical dynamics (e.g. cluster size). This could in principle be realised in synthetic settings, in the design of adaptive mechanochemical materials from chemotactic gels \cite{Dayal2013} or synthetic motor-filament systems \cite{Sanchez2012,Ross2019}, and in the realisation of active computational fluids \cite{Woodhouse2017}.

Aside from possible synthetic realisations, the schema considered here -- the triggering of active clustering upon external stimulus -- resembles many of the signalling reactions in biological systems, where nanoscale complexes of signalling receptors form larger scale aggregates to set up robust signalling outputs. The best studied examples are at the cell surface, including integrin focal adhesions \cite{VanZanten2009,Kalappurakkal2019}, signalling T-cell and B-cell `microclusters' \cite{Kumari2014,Tolar2017}, and Eph-Ephrin clusters \cite{Chen2021}. More recently, it has been argued that transcriptional machinery forms `condensates' in response to stimuli \cite{Hnisz2017,Sabari2018,Cho2018}, suggesting the kind of mechano-chemical computational processes studied here.

While we have considered here the simplest cases, the scope of the computational abilities inherent to an active fluid can be broadened greatly by two extensions. First, transition rates between chemical states (e.g. molecular conformations) may depend on physical variables such as cluster size (as argued recently for Eph-Ephrin clusters \cite{Chen2021}) or local stress (e.g. mechanosensitive receptors). Second, the dynamics of cluster merger allows for their coalescence and the building up of larger structures, which can sequentially recruit signalling molecules of increasing elaboration. This provides a natural mechanics for the \textit{concatenation} of computations.

We end with two general remarks. First, the non-reciprocal mechanochemical underpinning of active computational modules ensure that they respond, in terms of their spatial localisation, flows, and internal transition rates, to spatially organised stress distributions in the cell and vice versa, leading to long-range, \textit{substrate-mediated} allosteric propagation \cite{Bray1995,Phillips2020} with memory. Second, the computational capabilities of an active medium point to a notion of chemical multiplexing, where the same set of components can be brought together to produce several distinct computational outcomes.

KH thanks the James S McDonnell Foundation for support via a postdoctoral fellowship in Understanding Dynamic \& Multi-scale Systems, and the Institute for Biophysical Dynamics at the University of Chicago for support via a Yen Postdoctoral Fellowship. SR and MR acknowledge J.C. Bose Fellowship from DST-SERB (India). MR acknowledges support from the Department of Atomic Energy (India), under project no.\,RTI4006, and the Simons Foundation (Grant No.\,287975). 

\bibliography{computation}

\end{document}



\title{ {\Large Robust Molecular Computation by Active Mechanics} \\ {\normalsize SI} }

\author{K. Husain, S. Ramaswamy and M. Rao}

\maketitle 

\section{Dynamics of size and composition of an active cluster}


As described in the main text, the dynamics of concentration of molecular species $\alpha$
\be \label{eq:conc}
\f{\p}{\p t}c_{\a} = -\grad \cdot \l c_\a \v \r + M_\a \nabla^2 \mu_\a +
\sum_{\beta \neq \alpha} \, k_{\beta \to \a}(\{c_\a \},\boldsymbol \sigma) c_{\beta} - k_{\a \to \beta}(\{c_\a \},\boldsymbol \sigma)c_{\a} 
\ee
consists of advection by the gel velocity $\mathbf{v}$, diffusion down gradients of the chemical potentials $\mu_{\a}$, and reactions governed by the conversion rates $k_{\a \to \beta}$. These last could depend on the local concentrations of chemical species $\{c_\a\}$ (if, for instance, one species acts as an enzyme) or on the local mechanical stress, but we shall ignore such possible dependences for simplicity.


The gel velocity $\mathbf{v}$ is governed by mechanical stresses $\boldsymbol \sigma$:
\bea
 \label{eq:velphi}
    && \gamma \v =   \eta \nabla^2 \v +  \nabla  \cdot \boldsymbol \sigma \nn \\
    && \boldsymbol\sigma =   \boldsymbol\sigma_0 - \zeta \Delta\mu \, \mathcal{S} \l \sum_\a \, c_\a \, \beta_\a \r \,\mathbb{I}
 \eea
where $\gamma$ and $\eta$ are friction and viscosity, $\beta_\a$ is positive, negative or zero for activators, inhibitors or passive components respectively, and $\mathcal{S}$ is an odd sigmoidal function, vanishing at the origin and saturating at a positive (negative) value for large positive (negative) argument. In the main text, we present the stress with $\mathcal{S}$ linearised around small argument for clarity.

\indent  Generically, for active systems driven by contractile stresses, the dynamics represented by the first term in Eq.~\ref{eq:conc}, supplemented with Eq.~\ref{eq:velphi}, consists of compact clusters that can move and merge with each other. The composition of a cluster is determined by coarse-graining over the stress correlation length scale $\xi$:
\be
\f{\p}{\p t} n_\a \equiv \f{\p}{\p t} \int_{\xi} d{\bf r} \, c_\a = \underbrace{- \int_{\xi} d{\bf r} \, \grad \cdot (c_\a \v) + M_\a \nabla^2 \mu_\a}_{\text{physical flows}} + \overbrace{\sum_{\beta \neq \alpha} \, k_{\beta \to \a} n_{\beta} - k_{\a \to \beta} n_{\a} }^{\text{chemical dynamics}}
\ee

\vspace{.2in}

\indent \textbf{Master equation dynamics} These dynamics -- of clusters that merge, grow, and dissipate -- suggests a description in terms of aggregation and fragmentation dynamics. In this picture, the central object is the probability that a cluster has composition $P(\{ n_\a \})$, which is the (joint) probability that a cluster has $n_1$, $n_2$, ...,$n_\a$ particles of type 1, 2, ..., $\a$. In the main text, this is written as $P(\ldots, n_\a,\ldots)$ for clarity. In what follows, we shall consider only a single isolated cluster that grows and shrinks by the addition of single particles. In this limit, $n_\a$ can only change in two ways -- gain and loss of species $\a$ from the cluster, and interconversion of species $\a \to \beta$, which leads to a master equation of the form:

\be\label{eq:totalmaster}
\f{\p}{\p t}P(\{n_\a\}) = \underbrace{\mbf{G}(\sigma)\circ P(\{n_\a\})}_{\substack{\text{gain and loss} \\ \text{(physical flows)}}} + \underbrace{\mbf{C}\circ P(\{n_\a\})}_{\substack{\text{interconversion} \\ \text{(chemical dynamics)}}}
\ee
\indent Here, the operators ${\bf G}$ and ${\bf C}$ encode the physical dynamics of gain and loss, and the chemical dynamics of species interconversion, respectively. Explicitly, they can be succinctly written using the raising and lowering operators $\hat{a}_\a^{+}$ and $\hat{a}_\a^{-}$, defined by $\hat{a}_\a^{-}f(\{n_\a\}) = f(\{ n_\a - 1\})$ and $\hat{a}_\a^{+}f(\{n_\a\}) = f(\{ n_\a + 1\})$. The interconversion term follows directly from the rates $k_{\a \to \beta}$:
\be
\mbf{C} = \sum_{\alpha} \sum_{\beta \neq \alpha} k_{\beta \to \a} \l \hat{a}_{\beta}^{+}\hat{a}_{\a}^{-} - 1 \r n_{\beta} + k_{\a \to \beta} \l \hat{a}_{\beta}^{-}\hat{a}_{\a}^{+} - 1 \r n_\a
\ee
\noindent while the gain and loss terms take the form:
\be
\mbf{G}(\sigma) = \sum_\a \l \hat{a}_\a^{-} - 1 \r k_{+,\,\a}(\{n_\a\}) + \l \hat{a}_\a^{+} - 1 \r \, n_\a \, k_{\a}(\{n_\a\})
\ee
\indent As written, the gain and loss rates ($k_{+,\,\a}$, $k_{\a}$) are generic -- depending on the cluster state $\{ n_\a \}$ as well as on the species $\a$. Physically, however, they arise from the physical flows in Eq.~\ref{eq:velphi}, and so we will momentarily place constraints on them. For the moment, however, we will allow them to be arbitrary functions of composition $\{ n_\a \}$, but the same for all $\a$ (i.e. $k_{+,\,\a}(\{n_\a\}) = k_+(\{n_\a\})$, $k_{\a}(\{n_\a\}) = k(\{n_\a\})$)

\vspace{.3in}

\indent \textbf{Dynamics of cluster size} We now obtain, from Eq.~\ref{eq:totalmaster}, a master equation for the overall size of the cluster, defined as the total number of particles:
\be
n = \sum_\a \, n_\a
\ee
\noindent in the limit in which the chemical dynamics of interconversion -- represented by the operator ${\bf C}$ -- are fast. To do so, we first split $P(\{n_\a\})$ into a conditional and a marginal distribution,

\be
P(\{n_\a\}, n) = P(\{ n_\a \} \vert n) \, P_n
\ee
\noindent where $P(\{ n_\a \} \vert n)$ is the probability that the cluster has composition $\{n_\a\}$ given that the total cluster size is $n$ (naturally, $n = \sum_\a n_\a$), and $P_n$ is the (marginal) probability that the cluster size is $n$. That the dynamics of interconversion is fast implies that $P(\{ n_\a \} \vert n)$ satisfies the steady-state condition,
\be\label{eq:chemicalsteadystate}
{\bf C} \circ P(\{ n_\a \} \vert n) = 0
\ee
\indent To get an equation for $P_n$, we sum Eq.~\ref{eq:totalmaster} over $n_\a$ subject to $\sum_\a n_\a = n$. Using Eq.~\ref{eq:chemicalsteadystate} to eliminate the ${\bf C} \circ P$ term, and using $\langle \circ \rangle_n$ to denote an average over the conditional distribution $P(\{ n_\a \} \vert n)$, we obtain:
\be
\f{\p}{\p t} P_n = \left\langle \sum_\a \l \hat{a}_\a^- - 1 \r k_+ \l \{ n_\a \} \r \right\rangle_n P_n + \left\langle \sum_\a \l \hat{a}_\a^{+} - 1 \r \, n_\a \, k(\{n_\a\}) \right\rangle_n P_n 
\ee
\indent As $\sum a_\a^{\pm}$ accounts for all ways to go from $n$ particles to $n \pm 1$ particles, it is straightforward to verify that, for arbitrary functions $f$: 
\be
\left\langle \sum_\a \hat{a}_\a^{\pm} \, f(\{ n_\a \}) \right\rangle_n = M \, b^{\pm} \, \left\langle f(\{ n_\a \}) \right\rangle_n
\ee
\noindent where $b^{\pm}$ are operators that raise and lower $n$, and $M = \sum_\a 1$ is the total number of species. Absorbing $M$ into the rates $k_+$ and $k$,
\be\label{eq:finalmasterequation}
\f{\p}{\p t} P_n(t) = \l \hat{b}^- - 1 \r \langle k_+ \rangle_n \, P_n +  \l \hat{b}^+ - 1 \r n \, \langle k \rangle_n \, P_n
\ee
\indent That is, the dynamics of $P_n$ are simply that of a birth-death process with state-dependent rates $\langle k_+ \rangle_n$ and $\langle k \rangle_n$. Once we have prescribed a form for the functions $k_{\pm}\l \{ n_\a \} \r$, Eq.~\ref{eq:finalmasterequation} and Eq.~\ref{eq:chemicalsteadystate} form a closed system of equations. We will now discuss the form of $k_{\pm}\l\{n_\a\}\r$.

\vspace{.3in}

\indent \textbf{Form of gain and loss rates} The rates of particle addition and removal from a cluster arise from the physical flows in Eq.~\ref{eq:velphi}. If the cluster contains `active' particles -- those capable of inducing or suppressing local contractile stresses $\sigma$ -- then the rates of gain and loss $k_+$ and $k_-$ are functions of the cluster composition $n_\a$. For simplicity, we will suppose that the contractile stresses are binary -- switching on and off depending on the composition $n_\a$. 

\textit{Case 1: active stresses `off'}: When the active stresses are `off', the gain and loss dynamics are driven by the chemical potential alone, such that the ratio of the rates of particles in and out of the cluster is given by 
\be
\f{k_{+,\,\a}}{k_{\a}} = \me^{-\beta\mu_\a}
\ee
\noindent where $\beta$ is an inverse (possibly effective) temperature. This represents relaxational dynamics towards the low background density of particles, which we assume to be small. Therefore, in the absence of active stresses, we will suppose that the rate of gain of particles, $k_+$, is $0$.

\textit{Case 2: active stresses `on'}: When the active stresses are `on', we have $\sigma_{ij} = -\zeta \Delta \mu \delta_{ij}$ which renormalises the chemical potential in Eqn.~\ref{eq:conc},
\be
\mu_\a^{\text{eff}} = \mu_\a + \zeta \Delta \mu 
\ee
\noindent Recall that $\zeta < 0$ as the stress is contractile.

Thereby, the flux of particles of species $\alpha$ into the cluster, $k_{+,\,\a}$, and the rate of particles of species $\alpha$ leaving the cluster, $k_{\a}$, obeys:
\be
\f{k_{+,\,\a}}{k_{\a}} = \me^{ - \beta \mu_\a^{\text{eff}}  } = \me^{-\beta\mu_\a} \,  \me^{-\beta \zeta \Delta \mu}
\ee
We shall suppose that $\mu_\a$ is small compared to $\zeta \Delta \mu$ and therefore drop the labels to simply arrive at $k_+/k \approx \me^{-\beta \zeta \Delta \mu}$. This argument constrains only the ratio $k_+/k$; for concreteness, we shall suppose that $k$ is simply a constant, setting units of time, and 
\be
k_+ = \lambda \equiv k \, \me^{-\beta \zeta \Delta \mu}
\ee


\section{Receptors as information processing units}

In what follows, we will use the master equation dynamics above to calculate probability distributions of cluster composition in different `environments'. The goal is to compute the mutual information between environmental state, represented by the variable $L$, and cluster composition $n_\a$. For concreteness, we will imagine that our cluster is composed of `receptors' that bind to different ligands; the presence and type of ligands specifies the environment. The kinetics of ligand binding are taken to be simple mass action, such that the probability that a receptor is bound to a ligand is
\be
\f{\left[ L \right]}{\left[ L \right] + K_D}
\ee
\noindent where $\left[ L \right]$ is the ligand concentration and $K_D$ is the dissociation constant. In practice, we will assume that $\left[ L \right]$ is either $0$ (no ligand present in the environment: $L = 0$) or $\left[ L \right] \gg K_D$ (saturating amounts of ligand present in the environment: $L = 0$).

The species index $\a$ is taken to be an internal conformational state of the receptor, which determines downstream signalling (and, potentially, activation or inhibition of active stresses). The transition rates between conformations depends on whether or not the receptor has a ligand bound; in the simplest case (analysed in Sec.~\ref{sec:ligrec}), the receptor has only two internal conformations, which we label by the binary variable $R = 0$, $1$. We shall refer to $R = 1$ as the `signalling' conformation. More complex cases are a straight-forward extension, and will be described later.

\vspace{.1in}

\textbf{Receptor fidelity} Instead of dealing directly with the transition rates, we parameterise the steady state of a single receptor by the conditional distribution $P(R \vert L)$:
\be
P(R \vert L) = 
  \begin{cases}
    L & \text{with probability $p$} \\
    1-L & \text{with probability $1-p$}
  \end{cases}
\ee
\noindent where, for simplicity, we have supposed that the probability of a `false positive' (i.e. $R = 1$ when $L = 0$) is equal to the probability of a `false negative' ($R = 0$ when $L = 1$). There is only a single parameter $p$: the `fidelity' of this molecular channel.

\indent This model for a ligand-receptor system corresponds to a binary symmetric channel, for which we may straightforwardly compute the mutual information between ligand presence and receptor activation as:
\be
I(R;L) = 1 + p\log p + (1-p)\log (1-p)
\ee
\noindent assuming a maximally uninformative environment, $P(L = 0) = P(L = 1) = 1/2$.

\indent If we think of a single receptor as a single information processing unit -- that is, a channel by which information about the environment is obtained -- then we see that the performance of this channel depends strongly on $p$. For instance, a single-channel fidelity of $p=0.75$ results in an information transmission capacity of only $I \approx 0.19$ bits. To therefore extract reliable information from a single channel, then, requires extremely high molecular fidelity.

\vspace{.1in}

\textbf{In what follows} we will show that by coupling receptor signalling to active stresses, and thereby cluster size, clusters can reliably and efficiently extract information from the environment. We will do so with three representative examples:
\begin{enumerate}
    \item Binary classification (Fig. 2 of the main text)
    \item Discrimination (Fig. 3 of the main text)
    \item Multiple-input computations (Fig. 4 of the main text)
\end{enumerate}

\section{Fig. 2: Ligand recognition}\label{sec:ligrec}





\indent Continuing with the example in the last section, how does the situation improve if, instead of relying on a single receptor, the output of a cluster of $n$ receptors is pooled? The number of receptors that are signalling, $m \leq n$, is binomially distributed:
\be \label{eq:binomialActivation}
P(m \vert L) = \binom{n}{m} q_L^m (1-q_L)^{n-m}
\ee
\noindent where $q_L$ is the probability that a single receptor is signalling: $q_L = p$ if $L = 1$ and $q_L = 1-p$ if $L = 0$.

\indent From $P(m \vert L)$ and $P(L) = 1/2$ we compute the mutual information $I(m;L)$ numerically. This is plotted in Fig. 2c in the main text (black points) as a function of cluster size $n$ (for $p = 0.6$) and in the inset of Fig. 2d (gray points) as a function of the single receptor fidelity $p$.

\indent We note two things. First, in agreement with our intuition, even a low-fidelity receptor can give rise to reliable ligand detection when the output of many receptors is pooled together. Second, achieving an appreciable overall fidelity requires many receptors. This resource `cost' can be quantified simply by the cluster size $n$, and so Fig 2c can be interpreted as a trade-off between information transmission and resource cost.

\vspace{.1in}

\textbf{Dynamic cluster} We now consider a cluster of receptors that regulates its own size by coupling to the active stresses discussed in the first section. As before, we shall compute the mutual information between the number of signalling receptors in a cluster, $m$, and ligand presence or absence $L$. To do so, we solve Eqs.~\ref{eq:finalmasterequation} numerically.

First, for a fixed $n$ and $L$ we compute $P(m \vert n,L)$ from Eq.~\ref{eq:binomialActivation}. We use this to compute the rate of addition of receptors to a cluster by assuming that the active stresses are present only when the majority of receptors in a cluster are signalling (this corresponds to a model in which the non-signalling receptor, $R = 0$, inhibits active stresses, while the signalling receptor, $R = 1$, activates them):
\be \label{eq:activerate}
k_+(n) = \lambda \sum_{m > n/2} P(m \vert n,L)
\ee
\indent With this expression, the master equation Eq.~\ref{eq:finalmasterequation} takes the form:
\be \label{eq:master}
\f{\p}{\p t} \begin{pmatrix} P_1 \\ P_2 \\ P3 \\ \vdots \end{pmatrix} = \underbrace{\begin{pmatrix}
-k_+(1) & 2\mu & 0 & 0 & \cdots \\
k_+(1) & -\mu + k_+(2) & 3\mu & 0 & \cdots \\
0 & k_+(2) & -\mu - k_+(3) & 4\mu & \cdots \\
\vdots & \vdots & \vdots & \vdots & \ddots \\
\end{pmatrix}}_{\bf M}
\begin{pmatrix} P_1 \\ P_2 \\ P3 \\ \vdots \end{pmatrix}
\ee
\indent We solve for the steady state occupancy $P_n$ as the vector spanning the null-space of the tridiagonal matrix ${\bf M}$. In practice, we do so numerically by truncating the master equation to $n = 300$.

\indent Finally, we use $P_n$ to compute the marginal distribution of signalling receptors:
\be
P(m \vert L) = \sum_n P(m \vert n, L) P_n
\ee
\noindent and thereby the mutual information $I(m;L)$. We also compute the average `cost' as the average size of the cluster, $\langle n \rangle$. In Fig. 2c of the main text, this is plotted as an orange point (for $p = 0.6$, $\lambda/k$ = 25); the mutual information as a function of $\lambda/k$ and of molecular fidelity $p$ is plotted in Fig. 2d.

\section{Fig. 3: Ligand discrimination}

In the living cell, recognition of an external stimulus typically initiates a sequence of downstream processes. Often, these downstream events consist of recruiting further signalling molecules to extract more information from the environment. To model this, we consider a hierarchically structured environment $L$, depicted in Fig. 3a of the main text as a binary decision tree. Correspondingly, our receptors have three states ($R = 0,1,2$) correspond to the three environmental states ($L = \phi$, triangle, square, respectively). 

\indent Once again, we parameterise the fidelity of the receptor by $0< p < 1$: when the ligand is not present, the receptor has probability $p$ of being correct and probability $(1-p)/2$ of erroneously reporting a square or triangle ligand present. When a triangle (square) ligand is present, the receptor faithfully reports triangle (square) with probability $p^2$, erroneously identifies the ligand as a square (triangle) with probability $p(1-p)$, or fails to detect any ligand at all with probability $(1-p)^2$. 

\indent Here, the state of the cluster is specified by $m_i$: the number of receptors in the state $R=i$, with $m_0 + m_1 + m_2 = n$ being the total number of receptors in the cluster. With $m = m_1 + m_2$, we recover the model of the last section. As before, we couple receptor activation (detection of a ligand, regardless of type) to contractile stresses, arriving again at the master equation Eq. \ref{eq:activerate}, \ref{eq:master}. The distribution of receptor activation types $P(\{m_0,m_1,m_2\} \vert n)$ is given by the multinomial distribution, generalising Eq. \ref{eq:binomialActivation}.

\indent We solve at steady state to obtain $P(n \vert L)$ and thereby $P(\{m_0,m_1,m_2\} \vert L)$. To simultaneously characterise ligand recognition and discrimination, we consider two separate mutual information measures. Defining $m_r = m_1 + m_2$, we compute the information associated with ligand recognition, regardless of ligand type:
\be \label{eq:mI_recog}
I_{\text{recog}}(m \equiv m_1 + m_2 ; L)
\ee
\indent By construction, this is identical to the mutual information measure computed in the last section. To analyse discriminatory power, we compute the information transmitted about ligand identity when one of the two ligands is present:
\be \label{eq:mI_discrim}
I_{\text{discrim}}(m_i ; L = \text{ triangle or square})
\ee
\indent A large $I_{\text{recog}}$ implies that the cluster can reliably distinguish between ligand presence and absence, independent of ligand type. Conversely, a large $I_{\text{discrim}}$ indicates a robust discriminatory ability.

\indent In Fig. 3c of the main text, we plot these two measures as a function of activity $\lambda/k$. Interestingly, while for the parameters chosen here an activity of $\lambda/k \sim 1$ is sufficient to saturate $I_{\text{recog}}$, robust discrimination is only achieved at higher values of activity. Thus, extracting more information from the environment requires a stronger coupling to active stresses.

\indent We then studied the \textit{dynamics} of information transfer, numerically solving the master equation $\p_t \vec{P} = {\bf M} \vec{P}$ in time as $\vec{P}(t) = \exp \l {\bf M} t \r \vec{P}(0)$. The initial condition $\vec{P}(0)$ is taken to be a monomer, $P_n = \delta_{n,0}$. The average cluster size $\langle n \rangle(t)$ and its standard deviation are plotted in Fig. 3d, as well as the mutual informations Eqs. \ref{eq:mI_recog} and \ref{eq:mI_discrim}. We see that $I_{\text{discrim}}$ rises after $I_{\text{recog}}$, showing that gaining information about deeper levels of a decision tree (Fig. 3a in the main text) requires a. longer times, and b. larger cluster sizes.

\section{Fig. 4: Multiple-input computations}

The previous two sections were examples of single input computations, in which the task is either recognition or discrimination. In this section, we will consider a different type of task -- multiple input computations. That is, the environment presents more than one signal (e.g. multiple ligands), on which some computation must be performed to respond to a specified combination of inputs.

\indent Consider the receptor architecture in Fig. 4a of the main text, which accepts two inputs $A$ and $B$ and performs a logical operation $G(A,B)$. The output of the operation is receptor activation ($R=1$) or quiescence ($R= 0$). Once again, we suppose that the receptor is characterised by a fidelity $p$, here corresponding to the correct recognition of each signal $A$ and $B$. The output of the receptor is therefore:

\be \label{eq:stochastic_logical_output}
R = 
\begin{cases}
G(A,B) \, \text{ with probability } p^2\\
G(\text{not}(A),B), \, \text{ with probability } p(1-p)\\
G(A,\text{not}(B)), \, \text{ with probability } p(1-p)\\
G(\text{not}(A),\text{not}(B)), \, \text{ with probability } (1-p)^2\\
\end{cases}
\ee

\indent The probability of activating a single receptor ($R=1$) is shown for the AND and XOR logical operations in the top-right of Fig. 4b,c (scaled relative to the maximal probability).

\indent Regardless of the choice of logical operation $G$, we suppose that receptor signalling ($R=1$) induces contractile stresses and recruits further receptors to the cluster. Armed with the stochastic rules of Eq. \ref{eq:stochastic_logical_output}, we solve the master equation Eq. \ref{eq:finalmasterequation} at steady state and compute the distribution of number of signalling receptors $m$ (e.g., those with $R = 1$). 

\indent In Fig. 4b,c we show the distributions $P(m)$ for each value of the inputs $A,B$, for the case of the AND and XOR operations, respectively. In each case, the active cluster correctly achieves states of low and high signalling output $m$ based on the desired truth table. To compare against the case of the single receptor, we took as metric for cluster activation the average $\langle m \rangle$. We show this on the lower-right of Fig. 4b,c. Despite arising from an inherently noisy molecular computation, the output of the cluster robustly reproduces the AND and XOR truth tables.